\documentclass[12pt]{article}
\usepackage{graphicx,textcomp}
\usepackage{amsmath,amsthm,amssymb}

\usepackage{scicite}

\usepackage{times}

\newenvironment{sciabstract}{%
\begin{quote} \bf}
{\end{quote}}

\newcounter{lastnote}

\title{Evidence for Water in the Rocky Debris\\ of a Disrupted Extrasolar Minor Planet}

\author
{J. Farihi$^{1,4\ast}$, B. T. G\"ansicke$^{2}$, D. Koester$^{3}$\\
\\
\normalsize{$^{1}$Institute of Astronomy, University of Cambridge, Cambridge CB3 0HA, UK}\\
\normalsize{$^{2}$Department of Physics, University of Warwick, Coventry CV5 7AL, UK}\\
\normalsize{$^{3}$Institut f\"ur Theoretische Physik und Astrophysik, University of Kiel, 24098 Kiel, Germany}\\
\normalsize{$^{4}$STFC Ernest Rutherford Fellow}\\
\\
\normalsize{$^\ast$To whom correspondence should be addressed; E-mail:  jfarihi@ast.cam.ac.uk}
}

\date{}

\begin{document} 


\baselineskip24pt


\maketitle 


\begin{sciabstract}

The existence of water in extrasolar planetary systems is of great interest as it constrains the potential for habitable 
planets and life.  Here, we report the identification of a circumstellar disk that resulted from the destruction of a water-rich
and rocky, extrasolar minor planet.  The parent body formed and evolved around a star somewhat more massive than the 
Sun, and the debris now closely orbits the white dwarf remnant of the star.  The stellar atmosphere is polluted with metals 
accreted from the disk, including oxygen in excess of that expected for oxide minerals, indicating the parent body was 
originally composed of 26\% water by mass.  This finding demonstrates that water-bearing planetesimals exist around 
A- and F-type stars that end their lives as white dwarfs.

\end{sciabstract}

The enormous recent progress in the discovery of exoplanetary systems provides a growing understanding of their 
frequency and nature, but is still limited in many respects.  There is now observational evidence of rocky exoplanets 
\cite{fre12,bat11}, yet transit depth plus radial velocity amplitude provide planet mass and radius (and hence density), while 
the bulk composition remains degenerate and model dependent.  Transit spectroscopy offers some information on giant 
exoplanet atmospheres \cite{sin11}, and planetesimal debris disks often reveal the signature of emitting dust and gas 
species \cite{lis12}, yet both techniques only scratch the surface of planets, asteroids, and comets.  Interestingly, white
dwarfs -- the Earth-sized embers of stars like the Sun -- offer a unique window onto terrestrial exoplanetary systems:
these stellar remnants can distill entire planetesimals into their constituent elements, thus providing the bulk chemical 
composition for the building blocks of solid exoplanets.  

Owing to high surface gravities, any atmospheric heavy elements sink rapidly as white dwarfs cool below 25\,000\,K 
\cite{koe09}, leaving behind only hydrogen and helium in their outermost layers, a prediction that is corroborated by 
observation \cite{zuc03}.  Those white dwarfs with rocky planetary system remnants can become contaminated by the 
accretion of small, but spectroscopically detectable, amounts of metals.\footnote{Astronomers use the term `metal' when 
referring to elements heavier than helium}  Heavy element absorption lines in cool white dwarfs are a telltale of external 
pollution that often imply either ongoing mass accretion rates above $10^8$\,g\,s$^{-1}$ \cite{koe06}, or large asteroid-sized 
masses of metals within the convection zone of the star \cite{far10a}.

In recent years, metal-rich dust \cite{jur09a,rea05} and gas \cite{gan06} disks, likely produced by the tidal disruption of 
a large asteroid \cite{deb12}, have been observed to be closely orbiting 30 cool white dwarfs [e.g.\ \cite{far12,xu12,gir12,
far10b,far09,jur07}] and provide a ready explanation for the metal absorption features seen in their atmospheres \cite{jur03}.  
The circumstellar material being gradually accreted by the white dwarf can be directly observed in the stellar photosphere to 
reveal its elemental abundances \cite{zuc07}.  These planetary system remnants offer empirical insight into the assembly 
and chemistry of terrestrial exoplanets that is unavailable for any exoplanet orbiting a main-sequence star.

Until now, no white dwarf has shown reliable evidence for the accretion of water-rich, rocky planetary material.  
Unambiguous signatures of icy asteroids at white dwarfs should include: 1) atmospheric metal pollution rich in refractory 
elements; 2) trace oxygen in excess of that expected for metal oxides; 3) circumstellar debris from which these elements 
are accreted, and, where applicable; 4) trace hydrogen (in a helium-dominated atmosphere) sufficient to account for the 
excess oxygen as H$_2$O.  Critically, the presence of a circumstellar disk signals that accretion is ongoing, identifies the 
source material, enables a confident, quantitative assessment of the accreted elemental abundances, and thus a calculation 
of the water fraction of the disrupted parent body.

The metal-enriched white dwarfs GD\,362 and GD\,16 both have circumstellar disks and relatively large, trace hydrogen 
abundances in helium-dominated atmospheres \cite{jur09b}, but as yet no available assessment of photospheric oxygen 
\cite{zuc07,koe05}.  These two stars have effective temperatures below 12\,000\,K and their trace hydrogen could potentially 
be the result of helium dredge-up in a previously hydrogen-rich atmosphere \cite{tre07}.  The warmer, metal-lined white dwarfs 
GD\,61 and GD\,378 have photospheric oxygen \cite{des08}, but the accretion history of GD\,378 is unconstrained (i.e.\ it does 
not have a detectable disk), and without this information, the atmospheric oxygen could be consistent with that contained in 
dry minerals common in the inner Solar System \cite{jur10}.  In the case of GD\,61, elemental abundance uncertainties have 
previously prevented a formally significant detection of oxygen excess \cite{far11}.

\begin{figure}
\begin{center}
\includegraphics[width=16cm]{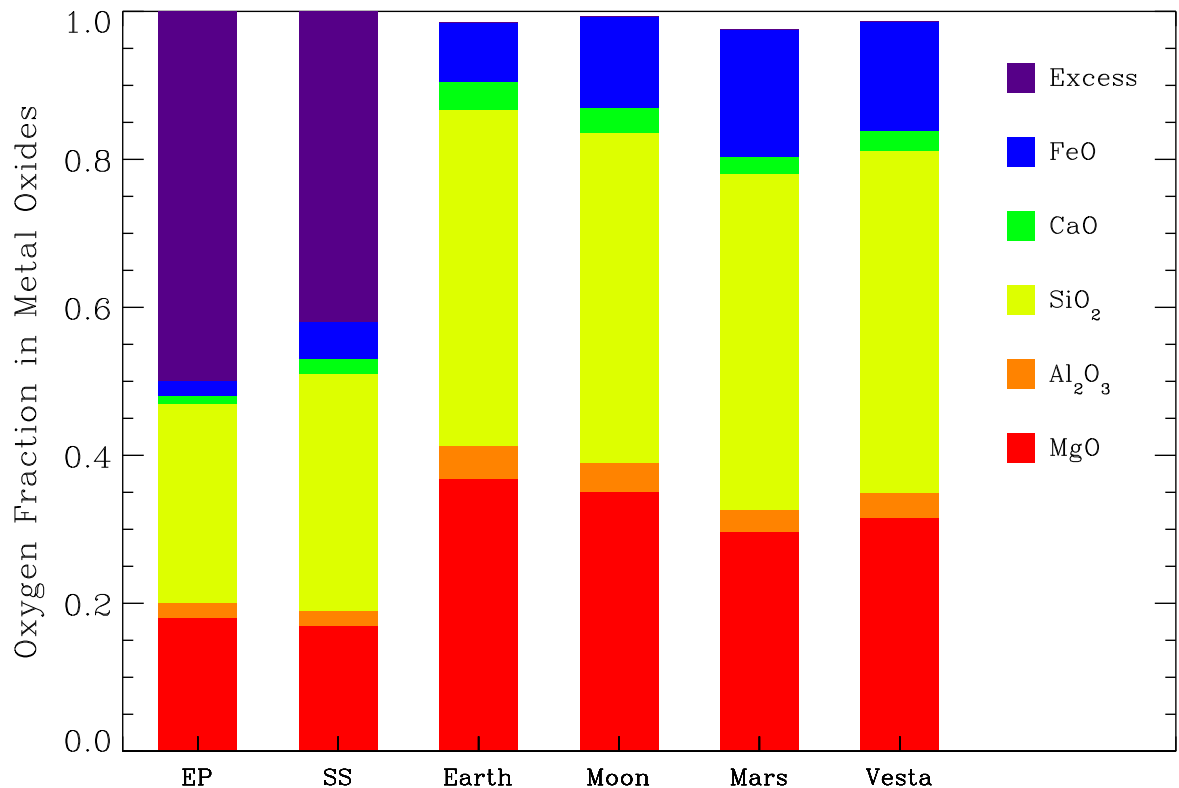}
\end{center}
\caption{The first two columns are the early phase (EP) and steady state (SS) fractions of oxygen carried by all the major 
rock-forming elements in GD\,61, assuming all iron is carried as FeO.  Additional columns show the oxide compositions of 
the bulk silicate (crust plus mantle) Earth, Mars, Moon, and Vesta \cite{vis13}.  Their totals do not reach 1.0 as trace oxides 
have been omitted.  The overall chemistry of GD\,61 is consistent with a body composed almost entirely of silicates, and
thus appears relatively mantle-like, but with significant water.  In contrast, the Earth is relatively water poor and contains
approximately 0.023\% H$_2$O (1.4$\times$10$^{24}$\,g).
\label{fig1}}
\end{figure}

We obtained ultraviolet spectroscopy with the Cosmic Origins Spectrograph (COS) on board the {\em Hubble Space 
Telescope} of the white dwarf GD\,61, and together with supporting ground-based observations, derived detections or limits 
for all the major rock-forming elements (O, Mg, Al, Si, Ca, Fe).  These data permit a confident evaluation of the total oxygen
fraction present in common silicates within the parent body of the infalling material, and we identify excess oxygen due 
to H$_2$O as follows.  1) The observed carbon deficiency indicates that this element has no impact on the total oxygen 
budget, even if every atom is delivered as CO$_2$.  2) The elements Mg, Al, Si, and Ca are assumed to be carried as MgO, 
Al$_2$O$_3$, SiO$_2$, and CaO at the measured or upper limit abundance.  3) The remaining oxygen exceeds that which 
can be bound in FeO, and the debris is interpreted to be water-rich.  We find oxygen in excess of that expected for anhydrous 
minerals in the material at an H$_2$O mass fraction of 0.26 (Table \ref{tbl1}, Fig. \ref{fig1}).

Because we have assumed the maximum allowed FeO, and some fraction of metallic iron is possible, the inferred water 
fraction of the debris is actually bound between 0.26 and 0.28.  Although this makes little difference in the case of GD\,61, 
where the parent body material appears distinctly mantle-like \cite{far11}, there are at least two cases where metallic iron 
is a major (and even dominant) mass carrier within the parent bodies of circumstellar debris observed at white dwarfs 
\cite{gan12}.  Overall, these data strongly suggest the material observed in and around polluted white dwarfs had an 
origin in relatively massive and differentiated planetary bodies.

We have assumed a steady state between accretion and diffusion in GD\,61.  However, a typical metal sinking timescale for 
this star is $10^5$\,yr, and thus the infalling disk material could potentially be in an early phase of accretion where material 
accumulates in the outer layers, prior to appreciable sinking \cite{far11}.  In this early phase scenario, the oxygen excess 
and water fraction would increase relative to those derived from the steady state assumption, and hence we confidently 
conclude that the debris around GD\,61 originated in a water-rich parent body.  Although the lifetimes of disks at white 
dwarfs are not robustly constrained, the best estimates imply that the chance of catching GD\,61 in an early phase is 
less than 1\% \cite{gir12,kle10,jur08,far08}.

The helium-rich nature of GD\,61 permits an assessment of its trace hydrogen content and total asteroid mass for a single 
parent body.  The total metal mass within the stellar convection zone is 1.3$\times$10$^{21}$\,g, and roughly equivalent 
to a 90\,km diameter asteroid.  However, because metals continuously sink, it is expected that the destroyed parent body 
was substantially more massive, unless the star is being observed shortly after the disruption event.  In contrast, hydrogen 
floats and accumulates, and thus places an upper limit on the total mass of accreted, water-rich debris.  If all the trace 
hydrogen were delivered as H$_2$O from a single planetesimal, the total accreted water mass would be 5.2$\times$10$
^{22}$\,g, and a 26\% H$_2$O mass fraction would imply a parent body mass of 2$\times$10$^{23}$\,g, which is similar 
to that of the main belt asteroid 4 Vesta \cite{rus12}.  

Based on these data, it appears that water in planetesimals can survive post-main sequence evolution.  One possibility is 
that solid or liquid water is retained beneath the surface of a sufficiently large ($d>100$\,km) parent body \cite{jur10}, and 
is thus protected from heating and vaporization by the outermost layers.  Upon shattering during a close approach with a 
white dwarf, any exposed water ice (and volatiles) should rapidly sublimate but will eventually fall onto the star -- the feeble 
luminosity of white dwarfs is incapable of removing even light gases by radiation pressure \cite{far08}.  Another possibility 
is that a significant mass of water is contained in hydrated minerals (e.g.\ phyllosilicates), as observed in main-belt asteroids 
via spectroscopy and inferred from the analysis of meteorites \cite{riv02}.  In this case, the H$_2$O equivalent is not removed 
until much higher temperatures and such water-bearing asteroids may remain essentially unaffected by the giant phases of 
the host star.

The white dwarf GD\,61 contains the unmistakable signature of a rocky minor planet analogous to Ceres in water content 
\cite{tho05}, and probably analogous to Vesta in mass.  The absence of detectable carbon indicates the parent body of the 
circumstellar debris was not an icy planetesimal analogous to comets, but instead similar in overall composition to asteroids 
in the outer main belt.  This exoplanetary system originated around an early A-type star that formed large planetesimals similar 
to those found in the inner Solar System and which are the building blocks for Earth and other terrestrial planets.

\bibliographystyle{Science}

\clearpage

\begin{table}
%\small
\caption{Oxide and Water Mass Fractions in the Planetary Debris at GD\,61\label{tbl1}}
\begin{center}	
\begin{tabular}{lrr}

\hline \hline

Oxygen Carrier		&Steady State		&Early Phase\\

\hline

CO$_2$			&$<$ 0.002		&$<$ 0.002\\
MgO				&0.17			&0.18\\
Al$_2$O$_3$		&$<$ 0.02			&$<$ 0.02\\
SiO$_2$			&0.32			&0.27\\
CaO				&0.02			&0.01\\
FeO$^a$			&0.05			&0.02\\

\hline

Excess			&0.42			&0.50\\

\hline

H$_2$O in debris:	&0.26			&0.33\\

\hline \hline

\end{tabular}	
\end{center}

{\em Note}. We adopt the steady state values which assume accretion-diffusion equilibrium.\\

$^a$ All iron is assumed to be contained in FeO, while some metallic Fe will modestly increase the excess oxygen.

\end{table}

\clearpage

\clearpage

\begin{center}
{\LARGE {\em Supporting Online Material for}\\ 
Evidence for Water in the Rocky Debris\\ 
of a Disrupted Extrasolar Minor Planet} 
\end{center}

\noindent
We describe here in detail the observations and analyses supporting the main paper, specifically the spectroscopy of 
the metal-enriched, white dwarf atmosphere and the analytical link to the elemental abundances of the infalling planetary 
debris.  

\section*{Summary of the Observations and Datasets}

GD\,61 exhibits infrared excess consistent with circumstellar dust orbiting within its Roche limit \cite{far11}, and bears the
unambiguous signature of debris accretion via its metal-polluted atmosphere.

\begin{figure}
\begin{center}
\includegraphics[width=16cm]{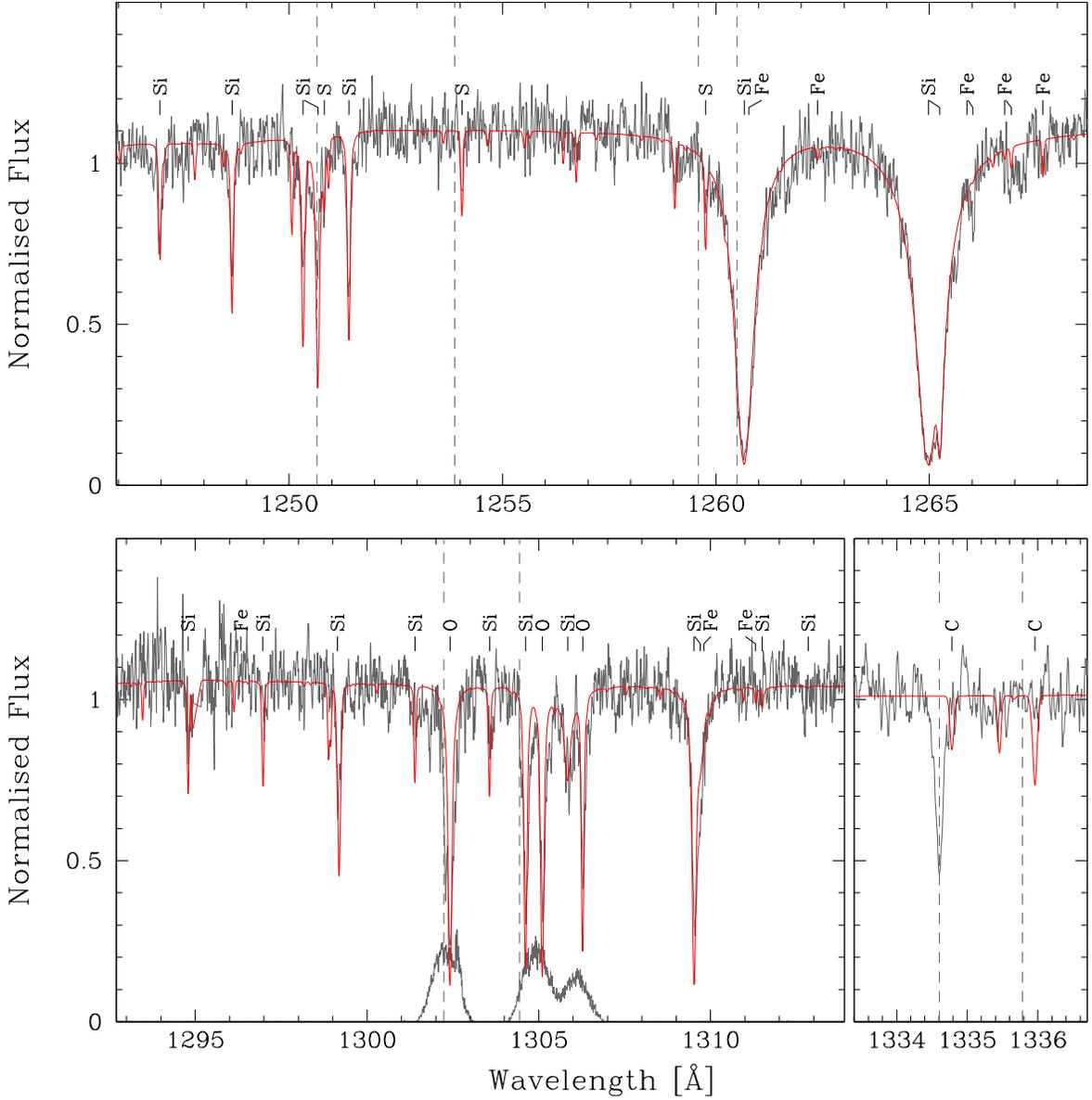}
\end{center}
\caption{The normalized COS spectra of GD\,61 (grey), together with the best fitting model spectra (red).   
Interstellar absorption features are indicated by vertical grey dashed lines, and are blueshifted with respect to the 
photospheric features by 40\,km\,s$^{-1}$.  Geocoronal airglow of O\,{\sc i} at 1302.2, 1304.9, and 1306.0\,\AA \ can 
contaminate COS spectra to some degree, and typical airglow line profiles are shown in the middle panel scaled to an 
arbitrary flux.
\label{fig2}}
\end{figure}

The white dwarf was observed with the Cosmic Origins Spectrograph (COS) during {\em Hubble Space Telescope} Cycle 
19 on 2012 January 28.  The ultraviolet spectra were obtained with a total exposure time of 1600\,s (split between two 
FP-POS positions) using the G130M grating and a central wavelength setting at 1291\,\AA, covering 1130$-$1435\,\AA \ 
at $R\approx$ 18\,000.  The COS data were processed and calibrated with {\sc calcos 2.15.6}, and are shown in Figure \ref{fig2}.  
Optical spectroscopy of GD\,61 was obtained on 2011 October 24 with the Keck II Telescope and the Echelle Spectrograph 
and Imager \cite[ESI]{she02} in echelle mode, effectively covering 3900$-$9200\,\AA \ at $R\approx$ 13\,000.  The spectra 
were obtained in a series of 16 exposures of 900\,s each, for a total exposure time of 4\,hr, and reduced using standard 
tasks in {\sc iraf}\footnote{IRAF is distributed by the National Optical Astronomy Observatories, which are operated by the 
Association of Universities for Research in Astronomy, Inc., under cooperative agreement with the National Science 
Foundation.}.  

\section*{Derivation of Photospheric and Debris Abundances}

Elemental abundances for GD\,61 were derived from the COS and ESI data by fitting white dwarf atmospheric models 
\cite{koe10} to the observed spectra.  For these calculations, $T_{\rm eff}=$ 17\,280\,K and $\log\,g=$ 8.20 are adopted, 
based on a published analysis of low-resolution optical spectra \cite{des08}.  The resulting photospheric abundances and 
upper limits are listed in Table \ref{tbl2} together with previous measurements from the {\em Far Ultraviolet Spectroscopic 
Explorer} \cite[FUSE]{des08} and Keck I HIRES \cite{far11}.  Notably, all heavy element abundances agree well, despite 
being derived using separate instruments and with multiple absorption lines across distinct wavelength regimes.

The transformation between the heavy element abundances in the white dwarf atmosphere and those within the infalling 
planetary debris are calculated assuming a steady state balance between accretion and diffusion.  An early (or build-up) 
phase of accretion is theoretically possible in GD\,61, but this is unlikely (see main paper).  Importantly, in this case an 
early phase would imply a larger oxygen excess and H$_2$O fraction, and therefore the more conservative, and most
probable, assumption is made.

For white dwarfs with significant convection zones like GD\,61, the atmospheric mass fraction $X_{\rm z}$ of heavy 
element z is related to its accretion rate $\dot M_{\rm z}$ via
\begin{equation}
\dot M_{\rm z} = X_{\rm z} M_{\rm cvz} / t_{\rm z}
\end{equation}
where $t_{\rm z}$ is the sinking timescale for the element and $M_{\rm cvz}$ is the mass of the stellar convection zone.  
The mass fraction is determined from the model atmosphere fits and the sinking timescale is known from white dwarf 
diffusion calculations \cite{koe09}.  In essence, Equation S1 states that the accretion rate of element z equals its rate 
of depletion as it settles below the mixing layer.  The ratio of two heavy elements within the debris (and hence parent 
body) is either the ratio of their respective accretion rates in the steady state, or the ratio of their atmospheric mass 
fractions in the early phase, and related by
\begin{equation}
\frac{\dot M_{\rm z1}}{\dot M_{\rm z2}} = \frac{X_{\rm z1}}{X_{\rm z2}} \times \frac{t_{\rm z2}}{t_{\rm z1}}
\end{equation}

Table \ref{tbl3} lists the relevant quantities of GD\,61 for the key elements that determine the total oxygen budget of the debris.  
The steady state metal abundances relative to oxygen are taken from the fourth column.  The sinking timescales for GD\,61 
have been updated following a correction in the theoretical calculations\footnote{http://www1.astrophysik.uni-kiel.de/\texttildelow{}koester/astrophysics/astrophysics.html}, and they are somewhat different than those presented in a previous analysis \cite{far11}.  
Notably, this correction has strengthened the case for an oxygen excess in GD\,61.

\section*{Evaluation of Oxygen Excess and Uncertainties}

The method for calculating the overall oxygen budget is as follows.  We begin with the columns in Table \ref{tbl3}, and in 
particular we identify the total oxygen budget with: 1) its mass accretion rate for the steady state or 2) its mass within the 
stellar convection zone for the early phase.  We calculate the fraction of oxygen that can be absorbed as CO$_2$ based 
on the upper limit for carbon, and subtract this from the total available.  Next, we perform a similar calculation for the mass 
of oxygen in MgO, Al$_2$O$_3$, SiO$_2$, CaO, and FeO based on their detections or upper limits, again subtracting these 
from the budget.  After accounting for all the major oxygen carriers, any remaining mass is considered excess.

The collective data for GD\,61 are robust and comprehensive, comprising four instruments with each probing distinct 
wavelength regions and containing multiple transitions for each element from the far-ultraviolet to the red optical region.  
The uncertainties in the metal abundances of this white dwarf are given as $3\sigma$ adopted values in the last column 
of Table \ref{tbl2}.  Using a brute force approach, all 128 possible combinations of abundance values are calculated for C, O, 
Mg, Al, Si, Ca, Fe where the abundances $N({\rm X})/N({\rm He})$ take on each of the values $x\pm\delta x$.  Evaluating 
all possible permutations, the dispersion in the resulting oxygen excess values (0.068) results in a $6.1\sigma$ confidence 
for the case of steady state accretion.

\clearpage

\begin{table}
\begin{center}
\caption{Elemental Abundances $N({\rm X})/N({\rm He})$ in GD\,61\label{tbl2}}
\vskip 15 pt
\begin{tabular}{ccccccc}

\hline

			&\multicolumn{2}{c}{Ultraviolet} 		&&\multicolumn{2}{c}{Optical} 				&\\
 
\cline{2-3}
\cline{5-6}

Element		&COS			&FUSE			&&ESI			&HIRES			&Adopted\\

\hline \hline

Detections:	&				&				&&				&				&\\

H    	 		&$-$3.70 (0.10)  	&				&&$-$4.00 (0.10)	&$-$3.98 (0.10)		&$-$3.89 (0.15)\\
O    	 		&$-$6.00 (0.15)		&$-$5.80 (0.20)		&&$-$5.75 (0.20)	&				&$-$5.95 (0.13)\\ 
Mg   	 		&$-$6.50 (0.30)  	&				&&$-$6.74 (0.10)	&$-$6.65 (0.18)		&$-$6.69 (0.14)\\ 
Si   	 		&$-$6.82 (0.12)  	&$-$6.70 (0.20)		&&$-$6.85 (0.10)	&$-$6.85 (0.09)		&$-$6.82 (0.11)\\ 
S    	 		&$-$8.00 (0.20)		&				&&				&				&$-$8.00 (0.20)\\
Ca                 	&	 			&				&&$-$7.77 (0.06)	&$-$7.90 (0.19)		&$-$7.90 (0.19)\\
Fe   	 		&$-$7.60 (0.30)		&$-$7.60 (0.20)		&&				&				&$-$7.60 (0.20)\\
\hline

Upper limits:	&				&				&&				&				&\\

C			&$-$9.10			&$-$8.80			&&				&				&\\
N			&$-$8.00			&				&&				&				&\\
Na			&				&				&&$-$6.80		&				&\\
P			&$-$8.70			&				&&				&				&\\
Al			&$-$7.80			&				&&				&$-$7.20			&\\
Ti			&$-$8.60			&				&&				&				&\\
Sc			&$-$8.20			&				&&				&				&\\
Cr			&$-$8.00			&				&&				&				&\\
Fe   	 		&				&				&&				&$-$7.50			&\\
Ni			&$-$8.80			&				&&				&				&\\

\hline \hline

\end{tabular}	
\end{center}
\end{table}

\clearpage

\begin{table}
\begin{center}	
\caption{Atmospheric and Debris Properties for Key Trace Elements in GD\,61\label{tbl3}}
\vskip 15 pt
\begin{tabular}{ccrr}

\hline

			&				&Early Phase				&Steady State\\
Element		&$t_{\rm diff}$		&$X_{\rm z} M_{\rm cvz}$$^a$	&$\dot M_{\rm z}$\\
			&(10$^5$\,yr)		&(10$^{21}$g)				&(10$^8$\,g\,s$^{-1}$)\\

\hline \hline

H			&$\infty$			&5.755					&\\
C			&1.730			&$<$ 0.001				&$<0.001$\\
O			&1.706			&0.802					&1.489\\
Mg			&1.808			&0.222					&0.389\\
Al			&1.735			&$<$ 0.019				&$<$ 0.035\\
Si			&1.438			&0.190					&0.419\\
S			&0.952			&0.014					&0.048\\
Ca			&0.782			&0.023					&0.091\\
Fe			&0.855			&0.063					&0.232\\

\hline

Total	Z		&				&1.332					&2.704\\

\hline \hline

\end{tabular}	
\end{center}

{\em Note}.  The metal-to-metal ratios within the planetary debris for the early phase and steady state regimes are derived 
directly from the values in the third and fourth columns respectively.

\medskip
$^a$The third column is the mass of each element residing in the convection zone of GD\,61, and their total (excluding 
hydrogen) represents a minimum mass for the parent body due to the continual sinking of metals.

\end{table}

\end{document}